\documentstyle[aps,epsf,pre,multicol,floats]{revtex}
\begin{document}

\title{Defects in Lamellar Diblock Copolymers:\\
Chevron- and Omega-shaped Tilt Boundaries}
\author{Yoav Tsori and David Andelman}
\address{School of Physics and Astronomy,
Raymond and Beverly Sackler Faculty of Exact Sciences\\
Tel Aviv University, 69978 Ramat Aviv, Israel}
\author{M. Schick}
\address{Department of Physics, Box 351560\\ University
of Washington, Seattle, WA 98195-1560}

\date{October 25, 1999}

\maketitle

\begin{abstract}

The lamellar phase in diblock copolymer systems appears as a result of a
competition between molecular and entropic forces which select a
preferred periodicity of the lamellae. Grain boundaries are formed when two
grains of different orientations meet.
We investigate the case where the lamellae meet symmetrically with respect to
the interface. The form of the interface strongly depends on the angle,
 $\theta$, between the normals of the grains. When this angle is small,
the lamellae transform smoothly from
one orientation to the other, creating the chevron morphology.
As $\theta$ increases, a  gradual
transition is observed  to an omega morphology characterized by a protrusion
 of the lamellae along the interface
between the two phases. We present a theoretical approach to find these tilt
boundaries in two-dimensional systems, based on a Ginzburg-Landau expansion of
the free energy which describes  the appearance of lamellae. Close to the tips
at which lamellae from different grains meet, these lamellae are distorted. To
find this distortion for small angles, we use a  phase variation ansatz in
which one assumes that the wave vector of the bulk lamellar phase depends on
the distance from the interface. Minimization of the free energy gives an
expression for the order parameter $\phi(x,y)$. The results describe the
chevron morphology very well. For larger angles,  a different approach is used.
We linearize $\phi$ around its bulk value $\phi_L$ and expand the free energy
to second order in their difference. Minimization of the free energy results in
a linear fourth order differential equation for the distortion field, with
proper constraints, similar to the Mathieu equation.  The calculated monomer
profile and line tension agree qualitatively with
transmission electron microscope experiments, and with full numerical solution
of the same problem.
\end{abstract}

\pacs{PACS numbers 61.25.Hq, 83.70.Hq, 61.41.+e, 02.30.Jr}



\newpage
\section{Introduction}\label{intro}
The lamellar phase is one of the possible
phases with spatial modulations that
can be found in a wide variety of physical and chemical systems. These include
diblock copolymer melts,  mixtures of diblock and homopolymers,
aqueous solutions of lipids or surfactants, Langmuir monolayers, and magnetic
garnet films \cite{SAscience}. Modulated phases are the result of a competition
between forces, one of which prefers ordering characterized by a
non-zero wavenumber, while the other prefers a homogeneous (disordered)
state.
Below we shall employ the language appropriate to block-copolymers, but
our work applies equally to other systems.

We consider
diblock copolymer melts in which the two polymer blocks are
incompatible. This incompatibility is characterized by a positive Flory
parameter $\chi$. Because of the covalent chemical bond between the A and B
blocks, the system cannot undergo a true macrophase separation.
Instead it undergoes a {\em microphase separation}
characterized by A-- and B--rich
domains of a finite size. Various modulated phases
such as lamellar, hexagonal, and cubic are observed
\cite{NAS,GT,Hashimoto,Leibler,FH,MB}
depending upon the $\chi$ parameter as well as on the relative lengths
of A and B
blocks. In most cases, the $\chi$ parameter depends inversely on the
temperature. Hence, the lamellar and other modulated phases will not be stable
at high temperatures and the polymer melt will be in a disordered state.

The equilibrium behavior of diblock copolymers in the bulk is by now well
understood \cite{Leibler,FH,MB}. Single domain bulk phases, however, are
rarely observed in experiments because it is extremely difficult to completely
anneal defects. In most cases, due to very slow dynamics and energy
barriers, the microstructural ordering is limited to finite-size domains (or
grains) separated by grain boundaries. These defects are very common
to block copolymer systems and are readily
observed in experiments \cite{GT,Hashimoto}.

Because these domain boundaries and defects are so abundant in polymer melts,
it is of interest to study their energetics and other characteristics. In this
paper we concentrate on the relatively simple situation of domain boundaries in
lamellar phases in which there is no twist between the two grains, only a tilt,
as is shown schematically in Fig.~1a. The system is translationally invariant
along the $z$-direction and can be described by its $x$-$y$ cross-section only,
reducing it to an effective two dimensional system. As can be seen in Fig.~1a,
the distance between lamellae along the grain boundary, $d_x$, is larger than
the lamellae spacing in bulk $d$, by a factor of $1/\cos(\theta/2)$. This
causes an increase in the local free energy density related to the grain
boundary.

The experiments of Gido and Thomas \cite{GT} and Hashimoto and
coworkers \cite{Hashimoto} show that the response of the system to this
increase in free energy depends strongly  on the tilt angle $\theta$ between
the grains. The so-called {\it chevron} morphology \cite{GT}
occurs when the angle is small. In such a
situation, lamellae  transform smoothly from one orientation to the other,
creating V-shaped tips. The rounding of these tips
reduces the interfacial area between
lamellae at the expense of introducing a curvature energy. For block co-polymers
the profile shape at the
tip is determined by the local relaxation of the stretched chains.
The chevron morphology is
shown (schematically) in Fig.~1b and again in Fig. 2a.
A gradual transition to an {\it omega-shaped} tip is
observed when $\theta $ is increased, as seen in Fig. 2b.
 This morphology is characterized by
protrusions of the lamellae along the interface between the two phases. The
protrusions can be understood as a different attempt of the system to reduce
the cost of the boundary, while still complying with the geometrical
constraints. Essentially the system tries to create a lamella similar to those
in the bulk but which is aligned along the interface itself.

The basic phenomenology of these grain boundaries was presented by Gido
and Thomas \cite{GT}. Netz, Andelman, and Schick\cite{NAS}
then considered the
phenomenon employing a Ginzburg-Landau free energy which was minimized numerically,
and obtained both the chevron and omega morphologies. Matsen\cite{matsen}
considered the block-copolymer system explicitly and employed
self-consistent field theory. Not only did this produce the chevron and
omega morphologies for small and intermediate angles, respectively,
but also a symmetry-broken omega for large enough angles. In the latter case,
the response (and shape) of lamellae
composed of one of the blocks differs from that of the other.
Such symmetry broken boundaries were indeed
observed by Gido and Thomas \cite{GT}.

In Sec. \ref{model} we adopt the Ginzburg-Landau free energy functional
employed earlier\cite{NAS}.
The advantages to using this
functional are its simplicity and generality, while retaining the essential
ingredients that capture the behavior of the system. In contrast to the
complete minimization of the free energy functional
which requires a numerical calculation, we shall employ here a
simple ansatz for the form of the grain boundary in order to obtain analytic
results. Recently, similar methods were employed
to obtain analytically the interface between
the lamellar and disordered phases of diblock copolymers \cite{VNAS}.
Our motivation is to demonstrate that the essence of these interesting
morphologies does not depend on strong segregation conditions or a large
number of Fourier components, and so should be observable in all systems
with modulated phases.

The chevron structure is obtained by using
the bulk lamellar phase solution with a constant amplitude,
but with varying wavevector.
Minimizing the free
energy subject to the proper geometrical constraints, we find an
equation for the wavevector. This is done in Sec. \ref{chevrons}. Beyond
the chevron regime ({\em i.e.} for large
inter-domain angles), this approach will not be adequate, because the amplitude
of modulations will have to vary as well. To this end we expand $\phi(x,y)$ around
the two  bulk lamellar phases, giving rise to an equation for a small
distortion field. Close to the interface, the sharp tips of the lamellar phase
(see Fig. 1a) are smoothed-out and the protrusion characteristic of
the omega morphology
appears. Far away from the interface, the disturbance vanishes and
the bulk lamellar phase is recovered,
as is shown in Sec. \ref{omegas}. In
Sec. \ref{degeneracy} we discuss some features of our method, the analogy
to, and the differences from the Schr\"{o}dinger equation for electrons in a
one-dimensional
periodic potential (known also as the Mathieu equation).
In Sec. \ref{results} we report our results,
and discuss them in Sec. \ref{summary}.

\section{the model}\label{model}

An order parameter $\psi({\bf R})=[\psi_A({\bf R})-
\psi_B({\bf R})]$ is defined as  the
difference in local
A and B monomer volume fractions.
We employ the following Ginzburg-Landau
free energy functional of this order parameter:

\begin{equation}
\frac{{\cal F}[\psi]}{k_BT}=\int\left[\frac{c_1}{2}\psi^2+\frac{c_2}{12}\psi^4
+\frac{c_3}{2}(\nabla \psi)^2+\frac{c_4}{2}(\nabla^2\psi)^2\right]{\rm d}^3{\bf R}
\end{equation}
where $k_B$ is the Boltzmann constant, and $T$ is the temperature.  With
$c_1$ and $c_2$ positive, the first two terms favor a uniform, disordered
state. The coefficient of the third term, $c_3$, is negative and
therefore induces the system to a modulated, ordered phase. The
Laplacian squared
term ensures that these modulations are not too large.  This type of
free energy functional (and some variants of it) has been successfully
used to describe the bulk phase of diblock copolymers \cite{Leibler,FH},
amphiphilic systems \cite{GS}, Langmuir films \cite{ABJ} and magnetic
(garnet) films \cite{GD}.

Assuming an infinite system that is non-uniform only along one
direction, one
can minimize this free energy to obtain a solution describing the
lamellar phase $\psi_L\propto \cos({\bf Q}\cdot{\bf R})$. One readily
finds that the transition to the lamellar phase first occurs at a
wavenumber $Q=(-c_3/2c_4)^{1/2}$. It is convenient, therefore, to
introduce the dimensionless position vector ${\bf r}$ via
\begin{equation}
{\bf r}\equiv \left(\frac{-c_3}{c_4}\right)^{1/2}{\bf R}
\end{equation}
and further to rescale the order parameter
\begin {equation}
\phi({\bf r})\equiv \left(\frac {c_2c_4}{c_3^2}\right)^{1/2}\psi({\bf R})
\end{equation}
and the free energy
\begin{equation}
F[\phi]\equiv \left(\frac{-c_2^2c_4}{c_3^5}\right)^{1/2}\frac{{\cal F}[\psi]}{k_BT}
\end{equation}
so that the dimensionless and rescaled free energy functional takes the
form
\begin{equation}
F[\phi]=\int \left[\frac{c_1c_4}{2c_3^2}\phi^2
+\frac{1}{12}\phi^4-\frac{1}{2} \left( \nabla
\phi \right)^2+\frac{1}{2}\left( \nabla ^2\phi \right)^2\right]{\rm d}^3{\bf r}
\end{equation}
The coefficient $c_1$ changes sign as the interaction strength increases or
temperature decreases.
In a polymer system, $c_1$ can be shown to be proportional to 
$1-\chi_f/\chi_f^*$, where $\chi_f$ is the 
Flory parameter\cite{flory}, and $\chi_f^*$ the value of this
parameter at which 
the coefficient $c_1$ passes through zero\cite{chi_s}.

To make contact with Ref. \cite{NAS}, we set the ratio of coefficients 
$c_1 c_4/c_3^2$ to be equal to $1-\chi$. 
For this particular choice of coefficients,
the free energy functional becomes
\begin{equation}
\label{F}
F[\phi]=\int \left[\frac{1}{2}\left(1-\chi\right)\phi^2
+\frac{1}{12}\phi^4-\frac{1}{2} \left( \nabla
\phi \right)^2+\frac{1}{2}\left( \nabla ^2\phi \right)^2\right]{\rm d}^3{\bf r}
\end{equation}

The bulk lamellar phase, with wavevector in the $x$ direction, is
described by

\begin{equation}\label{phiL1}
\phi_L=\phi_0+\phi_q\cos(qx)
\end{equation}
with $\phi_0$ the average volume fraction difference,
$q\equiv Q(-c_3/c_4)^{-1/2}=1/\sqrt{2}$ the optimal
wavevector, and $\phi_q=2\sqrt{\chi-\phi_0^2-3/4}$ the amplitude of the
variations \cite{VNAS}. It can be shown \cite{Leibler,FH,MB}
that this lamellar phase is the thermodynamical
stable phase in a range of the system parameters: $\phi_0$ and $\chi$.
For block copolymers, $\phi_0$ is proportional to the difference in the
average volume fractions of $A$ and $B$ monomers.

Upon substitution of $\phi_L$ into Eq. \ref{F}, one obtains the
free energy per unit volume of the bulk lamellar phase
\begin{equation}
\frac{F[\phi_L]}{V}=
\frac{1}{2}\left[(1-\chi)\phi_0^2+
\frac{\phi_0^4}{6}\right]-\frac{1}{2}\left[\chi- \frac{3}{4}-\phi_0^2\right]^2
\end{equation}
where $V=\int {\rm d}^3{\bf r}$ is the rescaled, dimensionless, volume of the system.

Some remarks are now in order.
First, as was mentioned in the introduction, the free energy (\ref{F}) can have
other non-lamellar modulated solutions \cite{NAS,spontak}. We will not
consider them in this paper since our aim is to study defects inside {\em
 lamellar} phases. Second,
the validity of a single optimal mode can be justified
in the weak segregation limit (i.e., near a critical point or a weak
first-order transition).
Far from the critical point higher harmonics are needed to describe
the optimal lamellar phase \cite{NAS,matsen}. 
In addition, very close to the critical point
corrections due to fluctuations are important\cite{BF,hamley}.

We now turn to the tilt-boundary problem, where two lamellar domains,
both lying parallel to the $x$-$y$ plane,
meet with an angle
$\theta$ between their normals. (See Fig.~1a.)
The $x$-axis is along the line
interface between the two lamellar domains. The
$y$-axis is perpendicular to it. In these variables the lamellae in the two
grains are  described by

\begin{equation}\label{phiL2}
\phi_L=\phi_0+\phi_q\cos(q_xx\mp q_yy)
\end{equation}
where $q_x\equiv q\cos \left(\theta/2\right)$ and
$q_y\equiv q\sin \left(\theta/2
\right)$ are the components of the optimal wavevector, ${\bf q}=(q_x,q_y)$.
Their inverses
provide characteristic length scales in the $x$ and $y$ directions,
respectively. The $y<0$ half plane is a reflection through the $x$ axis of the
$y>0$ half plane, so it is sufficient to consider only
the upper half plane, $y>0$. The system is
periodic along the $x$ axis, with wavelength $d_x=2\pi/q_x=
2\pi/[q\cos(\theta/2)]$.

\section{Description of the chevrons}\label{chevrons}

For small tilt angles, the lamellae transform smoothly from one orientation to
the other, showing the chevron morphology. We will assume that for small enough
tilt angles, the only change of the functional form of the order parameter is
through the wavevector \cite{dGP,Newell}. The aim of this section is to
show that, in the chevron regime, the behavior of the system close to the
interface is, in essence, quite similar to that far
from the interface. The diagonal
lines in Fig.~1a
show each bulk lamellar phase in its respective half-plane, and the sharp
tips which result from their intersection. These sharp tips will be smoothed
out in the chevron morphology, Fig.~1b.

We use the following ansatz for the order parameter:

\begin{equation}\label{cont phase}
\phi_c({\bf r})=\phi_0+\phi_q\cos[ q_x x + q_x u(y)],
\end{equation}
where the amplitude $\phi_q$ is identical to that in the bulk solution
$\phi_L$. This choice is motivated by the fact that, in the chevron
morphology, the amplitude of the order parameter appear to be rather
constant, while it is the phase which changes smoothly from one grain to
the other.
The local direction and magnitude of the wavevector depends
on the distance $y$
from the $x$-axis. Far away from the interface, the lamellae must return to
their bulk orientation, implying

\begin{equation}
\label{bc}
\lim_{y\rightarrow\pm\infty}u(y)=\mp\left(q_y/q_x\right)y= \mp
\tan\left(\theta/2 \right)y
\end{equation}
and at the interface, the continuous function $u(y)$ satisfies

\begin{equation}
\lim_{y\rightarrow 0}u(y)=0
\end{equation}

Symmetry with respect to inversion across the $x$-axis means that

\begin{equation}
\left.\frac{\partial\phi_c}{\partial y}\right|_{y=0}
=-\phi_qq_xu^{\prime}(0)\sin\left(q_xx\right)=0
\end{equation}
which implies
\begin{equation}
 u^{\prime}(0)=0
\end{equation}

We insert the form (\ref{cont phase}) into the free energy functional
(\ref{F}).  The integration over $x$ and $z$ can be carried out. The remaining
integration over $y$ shows that $F[\phi_c]$  is
proportional to the dimensionless, rescaled, volume $V$ of the
system. This reflects the fact that
the order parameter profile approaches its bulk value far from
the grain boundary. Subtraction of the bulk free energy produces
a functional for the
free energy per unit area of the grain boundary.
After simple manipulation, one finds the
expression
\begin{eqnarray}
\label{surf}
\gamma&\equiv&\frac{F[\phi_c]-F[\phi_L]}{A}\nonumber \\
      &=&\frac{\phi_q^2}{4}\sin^2(\theta/2)\int_0^{\infty}dy\left[
\left(\frac{ds}{dy}\right)^2+ \frac{\sin^2(\theta/2)}{2}(1-s^2)^2\right],
\end{eqnarray}
where
\begin{equation}
s(y)\equiv-\frac{1}{\tan(\theta/2)}\frac{du}{dy}.
\end{equation}
 From the boundary condition,
Eq. \ref{bc}, one sees that far from the grain boundary
$s$ attains the value $\pm 1$ and that its derivative vanishes.
It is clear, therefore, that if $s$ approaches its limiting value
sufficiently quickly, the grain boundary free energy is finite.
This is indeed the case.

The
Euler-Lagrange equation which minimizes the grain boundary free energy
is

\begin{equation}
\frac{d^2s}{dy^2}+s(1-s^2)\sin^2(\theta/2)=0
\end{equation}
which has the solution

\begin{eqnarray}
s&=&\pm\tanh\left[\frac{1}{\sqrt{2}}\sin(\theta/2)y\right]\nonumber\\
\label{sprofile}
&=&\pm\tanh q_y y
\end{eqnarray}
and
\begin{equation}
u(y)=\mp\frac{1}{q_y}\tan\left(\theta/2\right)\ln\cosh\left(q_yy\right)
\end{equation}
We have thus found
the order
parameter profile following the initial ansatz (\ref{cont phase})

\begin{equation}\label{u_y}
\phi_c(x,y)=\phi_0+\phi_q\cos[q_xx-\ln\left(2\cosh(q_y y)\right)]
\end{equation}
The grain boundary free energy can now be obtained
by inserting the profile solution, Eq. (\ref{u_y}),
into Eq. (\ref{surf})

\begin{equation}\label{st_TB}
\gamma=\frac{2}{3}\phi_q^2 q_y^3 \sim \sin^3(\theta/2)
\end{equation}

The order parameter we have calculated describes the chevrons very well. The tips of
the V-shape structure are rounded-off, and far away from the interface the bulk
phase is restored. The values of the grain-boundary energy are close
to those  obtained from full numerical minimization of the free energy
functional (\ref{F}), see Ref.~\cite{NAS}. The  expression (\ref{st_TB}) for the
grain boundary energy
shows a $\theta^3$
scaling for small angles \cite{dGP}

The width of the grain boundary is the characteristic distance over
which the phase of the order parameter profile deviates from its bulk
value.
From the profile of Eq. \ref{u_y}, we see that this distance is
$1/q_y$ which, for
small grain boundary angles, varies as $1/\theta$,
in accord with well established results \cite{dGP}.

The deficiency of the above approach is that it does not give the
cross-over from the chevron  to the omega morphology.
For this end another approach will be used in the next section.

\section{describing the omegas}\label{omegas}

As the tilt angle $\theta$ is increased, the chevron structure is deformed more
and more. For large angles the lamellae protrude along the interface between
the grains, creating the $\Omega$-shaped structure. In this regime the polymer
profile is qualitatively altered, and calls for a different approach.

In this section, we
use the bulk-phase solution as a zeroth order
approximation and determine a correction to it; that is we write
\begin{equation}
\phi(x,y)=\phi_L(x,y)+\delta \phi(x,y)
\end{equation}
Of course our ansatz for the chevrons, which was motivated by the smooth
variation of phase which they display, can also be written in this form
with a particular choice of $\delta\phi$. Below we shall choose a different
$\delta\phi$ based on the observation that the
distortion of the omegas is well localized near the grain boundary.

After substitution of this form into Eq. (\ref{F})
the free energy can  be written as a sum of two parts:

\begin{equation}
F[\phi]=F_L+\gamma A
\end{equation}
where $F_L=F[\phi_L]$ is the bulk free energy, proportional
to the volume and $\gamma A$ is the grain boundary energy proportional
to the area of the boundary.
To second order in $\delta \phi$, the latter  is

\begin{eqnarray}\label{deltaF}
\gamma[\delta\phi] &=&\frac{1}{A}\int
\{[ ( 1-\chi) \phi _L+\frac 13\phi _L^3]\delta \phi
+\frac 12( 1-\chi +\phi _L^2) \delta \phi^2
+\nabla ^2\phi _L\nabla ^2\delta \phi + \\
&&+\frac 12( \nabla^2\delta \phi) ^2-
\nabla \phi _L\nabla \delta \phi -\frac 12
( \nabla \delta \phi) ^2 \}{\rm d}^3{\bf r}\nonumber
\end{eqnarray}
Since $\phi_L$ minimizes $F_L$, we need to find the function $\delta
\phi(x,y)$ that minimizes $\gamma$. This, in principle, is done via the
Euler-Lagrange equation.

The boundary conditions for $\phi(x,y)$ follows from the symmetry of the
grain boundaries,
and from the requirement that $\phi$ approach its bulk value away from the
interface:

\begin{eqnarray}
\left.\frac{\partial^n \phi}
{\partial y^n}\right|_{y=0}=0 \label{BC1}\\
\nonumber \\
\lim_{y \rightarrow \infty }{\phi \longrightarrow \phi_L}
\label{BC2}
\end{eqnarray}
In the above, $n$ is odd. These conditions impose boundary conditions on
$\delta\phi$, because $\phi_L$ is known, and $\phi=\phi_L+\delta\phi$.

The distortion field $\delta \phi(x,y)$ will be found based on
an ansatz. Let us evaluate the $y$-derivative of the bulk tilted lamellar
phase:

\begin{equation}
\left.\frac{\partial \phi_L}{\partial y}\right|_{y=0}
=-q_y\phi_q\sin\left(q_xx \right)
\end{equation}
Because the distortion of the omegas is well localized near the grain
boundary at $y=0$, it is therefore reasonable to assume
that $\phi$  has the following form:

\begin{equation}\label{ansatz}
\phi_{\Omega}(x,y)=\phi_L(x,y)+f(y)\sin\left(q_xx\right)
\end{equation}

The procedure we adopt is as follows.
This  ansatz for $\phi$ is inserted into the free
energy density and the integration over $x$ and $z$ is carried out. The
grain-boundary energy functional $\gamma$ then depends on the unknown amplitude
$f(y)$.
This is minimized by an
Euler-Lagrange equation, which results in a fourth order ordinary differential
equation for $f(y)$. The boundary conditions, Eqs.~(\ref{BC1})-(\ref{BC2}),
will translate
into boundary conditions on $f(y)$.
The solution of the Euler-Lagrange equation gives, in principle, everything
we want
to know about
the system: spatial distribution of the order parameter, line tension, etc.

We begin by putting our ansatz (\ref{ansatz}) into the expression
(\ref{deltaF}) for
$\gamma$. Integration over $x$ and $z$ and a little simplification yields

\begin{eqnarray}\label{DF}
\gamma[f] &=&2\int_0^{\infty}\left\{ \frac{1}{4}q_y^2\phi_qf\sin q_yy \right.\\
&&+\frac 14\left[1-\chi+\phi _0^2+\frac 14\phi _q^2
\left( 1+2\sin^2q_yy\right)
+q_x^4-q_x^2\right] f^2  \nonumber \\
&&+\frac 12\phi _qq_yf^{\prime}\cos q_yy-\frac
14f^{\prime 2}+\frac 12q^2\phi _qf^{\prime \prime }\sin q_yy \nonumber \\
&&-\left. \frac12q_x^2ff^{\prime \prime }+\frac 14\left( f^{\prime
\prime}\right) ^2\right\}{\rm d}y
\nonumber
\end{eqnarray}

The Euler-Lagrange equation for the function $f(y)$ is obtained by
minimizing (\ref{DF})

\begin{equation}\label{gov eqn}
\left[ A+C\cos\left(2q_yy \right] \right)f+Bf^{\prime\prime}+
f^{\prime\prime\prime\prime}=0
\end{equation}
where $A,B$ and $C$ depend on the parameters of the problem as follows:

$A(\theta)=1-\chi +\phi _0^2+\frac 12\phi _q^2+q_x^4-q_x^2$

$B(\theta)=2q_y^2$

$C=-\frac 14\phi _q^2.$

Note that the profile equation (\ref{gov eqn}) is linear in $f$ because the
free energy, Eq. (\ref{deltaF}) is second order in $\delta\phi$.
Equation (\ref{gov eqn}) 
is  similar to the Mathieu equation, which is the
Schr\"{o}dinger equation for an electron in a periodic (sinusoidal)
one-dimensional potential
as appears in many solid state physics problems.
The parameter $A$ plays the
role of the electron total
energy, and $C$ is the amplitude of the periodic potential.
Unlike the
solid-state case,
here the ``energy'' parameter $A$, the ``kinetic term'' parameter $B$
as well as the
periodicity of the ``potential'' term depend on the angle $\theta$, leading
to a more complex band structure as function of $\theta$.
Lastly, the
differential equation is of fourth, not second order.
Another useful observation is that
equation (\ref{gov eqn})
is invariant with respect to the spatial variable transformation
$y \rightarrow -y$. Therefore, the solutions can be
classified as symmetric and
asymmetric. As we will see, the symmetric and asymmetric solutions do not
satisfy separately
the boundary conditions, so a combination of them will be needed.

Since this equation is linear with periodic coefficients, a solution to it will
have the Bloch form. This is also known as the Floquet theorem \cite{MF}:

\begin{equation}
f\left( y\right) ={\rm e}^{ky}g\left( y\right)  \label{floquet form}
\end{equation}
where $g\left( y\right) $ is a periodic function with period $d_y =\pi /q_y$
which is the same periodicity as appears in Eq.~(\ref{gov eqn}).
Hence it is appropriate to write it as a Fourier series:

\begin{equation}
g\left( y\right) =\sum\limits_{n=-\infty }^\infty a_n{\rm e}^{2inq_yy}
\end{equation}
Substituting the Bloch form Eq.~(\ref{floquet
form}) into  Eq.~(\ref{gov eqn}), we get a sum of exponential terms.
Demanding that the coefficients in front of every exponent vanish, we obtain
the following recursion relation:

\begin{equation}
\left( A+B\left( k+2inq_y\right) ^2+\left( k+2inq_y\right) ^4\right) a_n+\frac
12C\left( a_{n-1}+a_{n+1}\right) =0  \label{recursion}
\end{equation}

The appearance of $a_{n-1}$ and $a_{n+1}$ is due to the
$\cos\left(2q_yy\right)$ term
in (\ref{gov eqn}).
At first glance, it seems that for every $k$, choosing ``initial values''
for the
coefficients $\{a_n\}$ gives a valid solution. However, a closer inspection shows
that for an arbitrary $k$ vector the series $a_n$ will diverge. Only a very
specific value of $k$ (eigenvalue)  will give a convergent series.

The method by which we find this value is as follows
(e.g., see Ref. \cite{MF}). Rewrite the recursion relations
(\ref{recursion}) as

\begin{equation}
\frac{a_n}{a_{n-1}}=\frac{-\frac 12C}{A+B\left( k+2inq_y\right) ^2+\left(
k+2inq_y\right) ^4+\frac{1}{2}Ca_{n+1}/a_n}  \label{ratios plus}
\end{equation}
for $n>0$, and similarly

\begin{equation}
\frac{a_n}{a_{n+1}}=\frac{-\frac 12C}{A+B\left(k+2inq_y\right)^2+\left(
k+2inq_y\right)^4+\frac 12Ca_{n-1}/a_n}  \label{ratios minus}
\end{equation}
for $n<0.$ For very large values of $n$, $a_{n+1}/a_n$ should be much
smaller than
one, so one can start from some $N\gg 1$, assuming
that $a_{N+1}/a_N\rightarrow 0$,
and get

\begin{equation}
\frac{a_N}{a_{N-1}}\approx \frac{-\frac 12C}{A+B\left( k+2iNq_y\right)
^2+\left( k+2iN _y\right) ^4}  \label{approx N}
\end{equation}
then iterating {\em backward }gives the ratios $\{a_n/a_{n-1}\}$ for $0<n \leq N$.
Carrying out  the same procedure for negative $n$'s one arrives at
the {\em stopping
condition}:

\begin{equation}
A+Bk^2+k^4=-\frac 12C\left( \frac{a_{-1}}{a_0}+\frac{a_1}{a_0}\right)
\label{eqn for k}
\end{equation}
from which $k$ is deduced, because the ratios $a_1/a_0$ and $a_{-1}/a_0$ are
already known from Eqs.~(\ref{ratios plus})-(\ref{ratios minus}). The iteration
scheme is to choose an initial guess for $k$, put it in (\ref{approx N}), and
use (\ref{ratios plus}) and (\ref {ratios minus}) successively for all $n\neq
0.$ When $n=0$ is reached, $k$ is calculated from (\ref{eqn for k}), and put
back in (\ref{approx N}). The process repeats until convergence is achieved.
From the required boundary conditions at infinity, only $k$'s such that ${\rm
Re}(k)
<0$ are acceptable, recalling that the Bloch form (\ref{floquet form}) 
contains an ${\rm e}^{ky}$ term.

It can be seen from (\ref{ratios plus}), (\ref{ratios minus}) that if a certain
value of $k$ gives a physical converging solution for
$y\rightarrow \infty$, then $k^*$, the complex
conjugate of $k$,
will also be a convergent solution, but with amplitudes $a_{-n}^{*}$.
Solutions with $-k$
and $-k^{*}$ are possible, too, but are discarded because for them
$f(y)$ diverges as
$y\rightarrow\infty$.
The functions with definite symmetry include both $k$ and $-k$ and hence
diverge
at infinity.
Consequently, by dropping the $-k$ solution
we choose a specific combination of the
symmetric
and asymmetric functions with respect to $y$. The function
$f$ is a combination of the two independent solutions, and since it is real,
it must be equal to:

\begin{equation}
f\left( y\right) ={\rm e}^{ky}\sum\limits_{n=-\infty }^\infty a_n{\rm e}^{2inq_yy}
+c.c. ~~~~~~~~~~~~~~~~~~~{y\ge 0}
\label{f1 of y}
\end{equation}

A choice of  $k$ such that ${\rm Re}(k)<0$
ensures that the disturbance will decay away far enough
from the interface, $y\rightarrow \infty$. The use of
a linearization scheme
leaves us with a linear ordinary differential equation, so that we lose the
ability to impose all boundary conditions, and forces us to use only the
first and
third derivatives of $f$, which by the use of Eqs.~
(\ref{phiL1}), (\ref{BC1}) and (\ref{ansatz})
are

\begin{eqnarray}
\left.\frac{\partial f }{\partial y}\right|_0 &=&q_y\phi _q  \label{bc1}
\\
\left.\frac{\partial ^3f}{\partial y^3}\right|_0 &=&-q_y^3\phi _q
\label{bc2}
\end{eqnarray}
In summary, the series $\{a_n\})$
can be determined from the recursion relations
(\ref{ratios minus}), (\ref{ratios plus}), while
the only remaining unknown is the
complex parameter $a_0$. The function $f(y)$ itself
depends on this series as well as on
the two boundary conditions.
Substitution of (\ref{f1 of y}) in the above boundary
conditions fully
determines both the phase and magnitude of $a_0$ and therefore sets
the complete solution to the problem.
\subsection{Band-gaps and degeneracy}\label{degeneracy}

In the usual Schr\"{o}dinger equation
for periodic potentials, one encounter energetic gaps in
the energy spectrum.
They occur whenever the $k$-vector
crosses the edges of a Brillouin zone. It should be expected
that similar phenomenon happens here, too.
Indeed, in our case it will happen whenever

\begin{equation}\label{deg-cond}
{\rm Im}(k)=mq_y
\end{equation}
for some integer $m$. For $k$ fulfilling (\ref{deg-cond})
there is a {\em degeneracy} in the two previously found solutions. It can be
seen by noticing that if (\ref{deg-cond}) holds, $k$ and $k^{*}$ differ by
$2imq_y$.
Therefore, the amplitudes $\{a_n\}$ corresponding to the eigenvalue
$k^{*}$ are the same
as those $\{a_{n-1}\}$ corresponding to $k$, and the two solutions are dependent.
If $m$ is even, the two solutions are  real. One can see this by using
$k=k_r+2inq_y$ ($n$ being an integer) and
noticing that
\begin{equation}\label{equiv}
{\rm e}^{\left(k_r+2inq_y\right)y}\sum\limits_{n=-\infty }^\infty
a_n{\rm e}^{2inq_yy}=
{\rm e}^{k_ry}\sum\limits_{n=-\infty }^\infty
a_{n-1}{\rm e}^{2inq_yy}
\end{equation}
Using the right term of
(\ref{equiv}) in the iteration process it is clear that
$a_n=\left(a_{-n}\right)^*$, and $f(y)$ is real.

In these band gaps there is a {\em splitting} of the real part of $k$; another
{\em independent} $k$-vector appears whose imaginary part is the same, but
which has a different real part. This is clearly seen in Figure 8a,
where the band gaps appear to be centered around $18^{\circ}$, $26^{\circ}$
 and $42^{\circ}$. The last gap occurs for $\theta>127^{\circ}$.

The two solutions constructed by the Bloch
form (\ref{floquet form}) are obviously independent. It should be noted that
the appearance of energy gaps is a  mathematical consequence which
does not introduce any singularity into the physical grain-boundary
energy. It is due
to our approximative linearization scheme, and is not expected to
occur if non-linearity would have been included.

\section{Chevrons and Omega-like Profiles}\label{results}
We discuss the results obtained by using the analytical
eigenvalue scheme detailed above. For small angles (small tilt)
the response of the system is weak.
Namely, the lamellae gradually change their orientation
and the profile is well described by the analytical chevron
form, Eq. (\ref{u_y}).
At higher tilt angles  the polymer
chains at the interface are stretched much more than their preferred
length.  Here our analytical scheme (Sec.~IV) gives rise to
profile shape quite different than those obtained in Sec.~III for the chevron.

Figure 2 shows contour plots of the order parameter profile $\phi_{\Omega}(x,y)$.
We have taken the average of the order parameter, $\phi_0$, to be zero.
The gray levels indicate the magnitude of the order parameter: black
regions corresponding
to the maximum values of $\phi_{\Omega}$ (rich in A polymer), while white to
its minimum value (rich in B polymer). The interaction is set to $\chi=1$. 
In Fig.~2a the chevron morphology is represented using the results of Sec. IV,
with tilt angle
$\theta=20^{\circ}$. The plot is identical when using the expressions of Sec. III.
A smooth changeover between the two lamellar phases is
observed. The chevron solution follows nicely the analytical form
of Eq. (\ref{u_y}).
For large angles ($\theta=130^{\circ}$), the omega
structure takes over as is shown in Fig.~2b,
with large protrusions of the lamellae at the interface. As is explained below,
for large
angles the excessive packing frustration of the chevrons cost  more energy
than the omegas. 
Chevron and omega morphologies
are also shown in Fig.~3 but
with $\chi=0.76$ much closer to its critical value of $3/4$. As one goes
away from the $y=0$ interface, undulations of $\phi_{\Omega}$ are
encountered. There are more of them as the order-to-disorder
transition (ODT) is approached, and the omega structure is more evident.

Figures 4 and 5 show the same plots, but with the lines of interface enhanced.
The regions in white, gray and black correspond to $\phi_{\Omega}< -0.2$,
$-0.2\le\phi_{\Omega}< 0.2$, and $\phi_{\Omega}>0.2$, respectively. The gray
marks the interface between the A and B rich regions. The equi-$\phi_{\Omega}$
lines clearly show the form of the interface in the chevron and omega
morphologies. Notice that as the  ODT is approached ($\chi=0.76$ in Fig.~5),
there are more undulations apparent on top of the bulk lamellar phase. In the
chevron regime, the A/B interfacial width remains almost uniform, while in the
omega regime it varies close to the kink \cite{matsen}. Fig.~4 shows quite
clearly that the leading and trailing edges of the lamellae are quite different
when the angle of the grain boundary is large. This is emphasized in Fig.~6,
for $\theta=130^{\circ}$ and $\chi=1.0$, the same conditions of the lower plot
of Fig. 4. The lines indicate where $\phi=0.5$.
 That the protrusion at the leading edge is much more pronounced
than at the trailing edge is quite reminiscent of the profiles seen in
experiment\cite{GT} with one exception; in experiment, only half of the
lamellae look this way, the other half hardly display protrusions at all; that
is, the symmetry between positive and negative order parameter domains is
broken.

Results for the grain-boundary energy $\gamma_{\rm TB}$ are shown
in Fig.~7.
The $\chi$ parameter is arbitrarily fixed to be $\chi=0.76$. We show the
grain-boundary energy calculated analytically by means of the methods of Sec.
\ref{chevrons} and which is valid for small $\theta$, as well as that
calculated in Sec. \ref{omegas}. The grain-boundary energy $\gamma_{\rm TB}$ is
an increasing function of the angle $\theta$. In the small angle regime the
grain-boundary energy obtained in Sec. \ref{chevrons} scales as $\gamma \sim
\theta ^3$. This scaling is also satisfied (to a good approximation) by the
solutions obtained in Sec. \ref{omegas}. To see this, the same data is plotted
in a different fashion in Fig.~8. In 8a, $\gamma^{1/3}$ is plotted as function
of $\theta$, while in 8b the data is plotted on a log-log plot. From both parts
of Fig.~8 we conclude that the solution obtained in Sec. \ref{omegas} gives a
power law with exponent of $2.91$. Accuracy of the leftmost point of the solid
curve in Fig.~8b is doubtful, due to poor convergence of the numerical
iteration scheme for small angles, $\theta<5^{\circ}$. As the tilt angle grows,
deviations from the $\theta ^3$  behavior become larger, and the omega morphology, with
its lower energy, appears gradually.

For intermediate and large
angles, our results are supported by a full
numerical solution of this problem \cite{NAS}.
Presumably, for small
angles there is agreement too. This needs to be further checked since
in Ref. \cite{NAS} the smallest angle was $\theta=28^{\circ}$.
Note that in contrast to the
full numerical solution in which the order parameter profile was obtained via a
functional minimization, here we employed numerical means only to  obtain the
value of the eigenvalue $k$, while the profile equation was solved
analytically.

In Fig.~9 we show the $k$-vectors found by
the use of the iteration scheme, as a function of tilt angle $\theta$, with
$\chi=1$ fixed. Outside of the band-gaps, $k$ and $k^{*}$ are valid solutions,
so for clarity only the $k$ with ${\rm
  Im}(k)>0$ is shown. Notice that ${\rm Re}(k)$ is always negative. The band-gaps are
clearly seen
on this graph as regions where ${\rm Re}(k)$ have two distinct
values (hollow circles). A check on the
imaginary part of $k$ in these regions reveals that it is an integer
multiple of $q_y$.

Our results agree well with experiment with the exception that we
do not obtain the symmetry-breaking transition of the omega
morphology\cite{matsen}. To do so, one must add to an ansatz for the order
parameter at least a term which varies as $2q_x x$, in addition to the
fundamental term varying as $q_xx$.

\section{Concluding remarks}\label{summary}

We have used a simple Ginzburg-Landau free energy functional to investigate the
profiles between lamellar phases of diblock copolymer. Our analytic results
give  good qualitative agreement with experiment in the weak segregation
regime, and with full numerical solution of the same free energy model. The
observed chevron morphology develops gradually into an omega morphology for
intermediate tilt angles.
For
small angles, the use of a periodic order parameter with constant amplitude
but varying wavevector  suffices to describe the
order parameter profile. This is well described by the chevron morphology.
For intermediate angles the change of the profile at
the interface deviates significantly from the bulk, and requires a different
treatment. The deviation  from the bulk lamellae was found, and gave
rise to the protrusion characterizing the omega morphology.
The symmetry breaking of this phase was not obtained.

We were able to calculate grain-boundary energies and to determine that
they scale as the cube of the angle \cite{NAS,dGP} for small angles.  As
the tilt angle grows into the intermediate regime, the profiles deviates
continuously from the chevron shape.  As the angle approaches $180^
{\circ}$, the energy must go to zero because the grain boundary
vanishes at that angle.  This does not occur in our analytic
derivation, since, for such large angles, the linearization assumption
is no longer valid.

Interesting extensions of the present work would describe
interfaces
of two perpendicular lamellar phases (so-called {\em T-junctions}),
 interfaces
between modulated phases of other symmetries, and the inclusion of a twist
instead of a tilt \cite{GT2,lubensky}.

\bigskip
\noindent
\acknowledgements We benefited from discussions with
Y. Cohen, S. Gido, R. Netz, P. Rosenau,
M. Schwartz and E. Thomas.
Partial support from the U.S.-Israel Binational Foundation (B.S.F.)  under grant
No. 98-00429, the National Science Foundation under Grant No. DMR
9876864, and the Israel Science Foundation founded by the
Israel Academy of Sciences and Humanities --- centers of Excellence
Program  is gratefully acknowledged.

\newpage

\begin{itemize}

\item{\bf Fig.~1:}
{A schematic drawing of the geometry of the system. In (a) the tilt angle
$\theta$ between the normal of the two lamellar phases is shown. The bulk
periodicity $d$ is smaller then the local periodicity $d_x=d/\cos(\theta/2)$
 at the interface,
$y=0$. In (b) a schematic drawing of a chevron morphology with rounded V-shaped
tips is shown.
}

\item{\bf Fig.~2:}
Black and white contour plot of the order parameter profile, as obtained from
the profile solutions
of Sec. \ref{omegas}. The grayness denotes
the value of the order parameter (A/B relative volume fraction).
Black domains are A-rich, white domains
are B-rich. The value of the interaction parameter is $\chi=1$ and
the average difference of monomer volume fractions is $\phi_0=0$. For the
small angle ($\theta=20^{\circ}$, top plot) chevron morphology appears, while
for $\theta=130^{\circ}$ (bottom plot) omega takes over.

\item{\bf Fig.~3:}
Same as Fig.~2,
but with $\chi=0.76$. Top plot is for $\theta=20^{\circ}$ and the
bottom plot is
for $\theta=130^{\circ}$. The system is closer to
the ODT than that of Fig.~2,
and the modulations of the lamellar thickness are more prominent.

\item{\bf Fig.~4:}
Black, white and gray contour plot, showing A-rich, B-rich and
interfacial regions, respectively. Black regions are A-rich
($\phi_{\Omega}>0.2$),
white regions are B-rich ($\phi_{\Omega}<0.2$), while
gray regions marks the interfacial region ($-0.2\le\phi_{\Omega}\le 0.2$).
Top plot is for
$\theta=20^{\circ}$ and in the bottom plot $\theta=130^{\circ}$. The
interaction is set to $\chi=1$.

\item{\bf Fig.~5:}
As in Fig.~4,
but with $\chi=0.76$. Top plot is for $\theta=20^{\circ}$ and the
bottom plot is for $\theta=130^{\circ}$.
Comparison with Fig.~4 shows that for fixed tilt angle $\theta$,
the omega structure and the modulations of the lamellar
thickness are more evident closer
to the ODT.

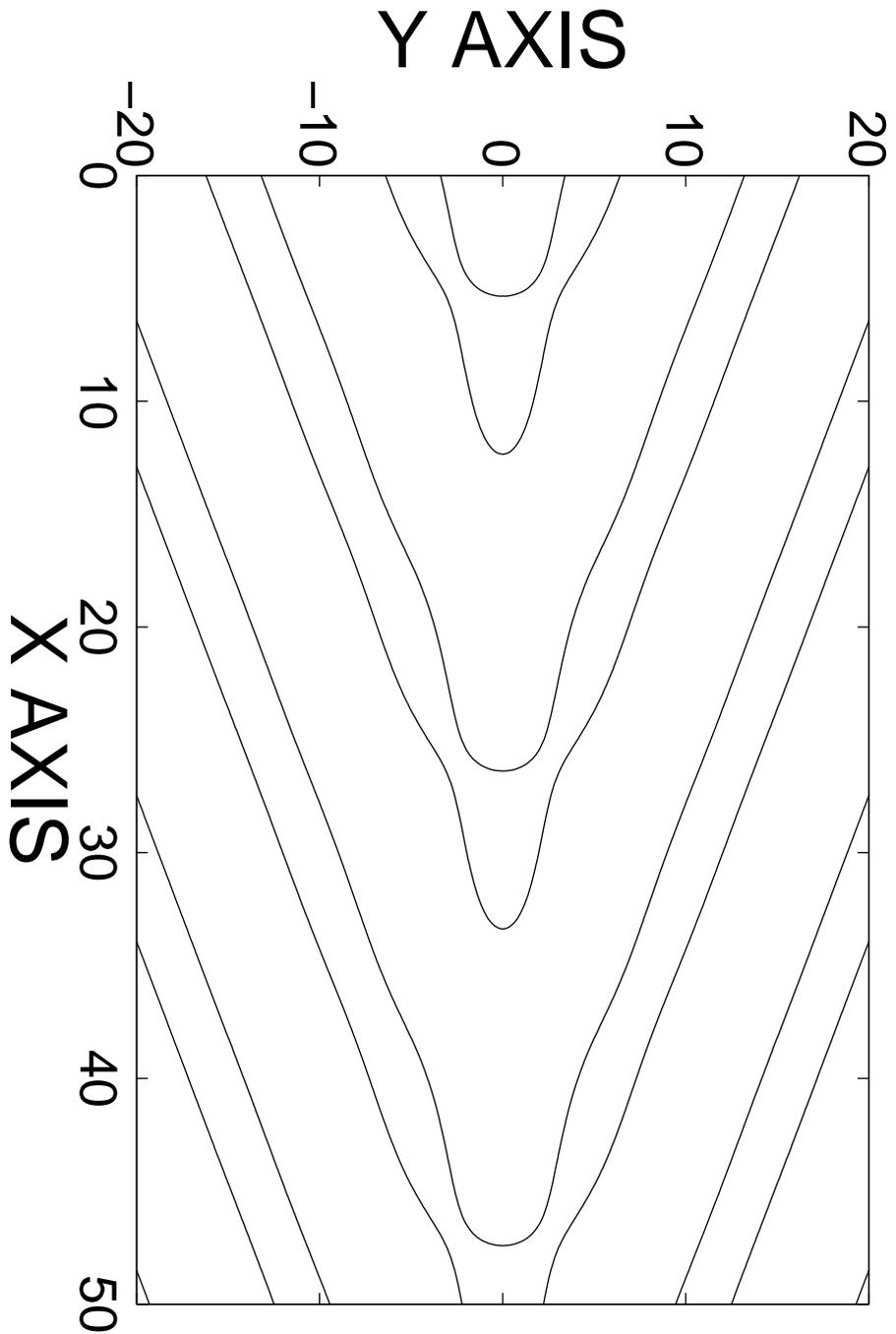
\item{\bf Fig.~6:} Plot of the equi-$\phi$ contour lines corresponding to
$\phi=0.5$, with the same conditions as in Fig.~4b. Although the profile
shown in Fig.~4 is
symmetric with respect to the interchange of A and B, clearly the leading and
trailing edges of the lamellae are not identical.

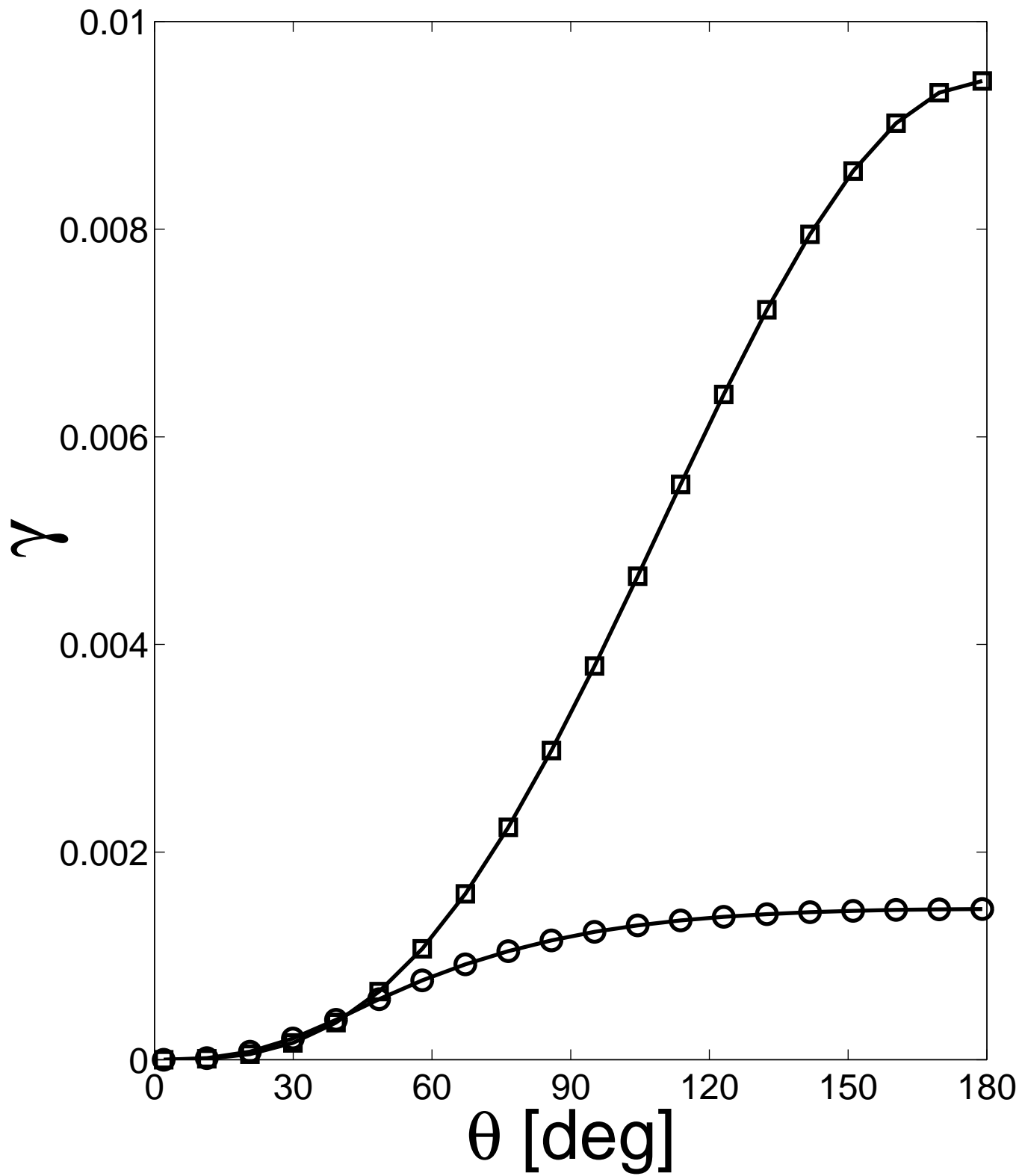
\item{\bf Fig.~7:}
Plot of the grain-boundary energy $\gamma$ as function of the tilt
angle $\theta$, for $\chi=0.76$. $\theta$ ranges from $0$ to $180^{\circ}$. The
 curve marked with rectangles shows the results obtained in Sec.
\ref{chevrons} for the chevron morphology,
with $\gamma=\frac23\phi_q^2q_y^3$. The curve with
circles shows the results of Sec. \ref{omegas}. Notice that the two
expressions are very similar, but at some intermediate
values of $\theta$ ($\theta \simeq 50^\circ$)
there is a crossover from the
chevron to the omega morphology. Evidently, for large tilt angles
the omegas cost less energy than the chevron.

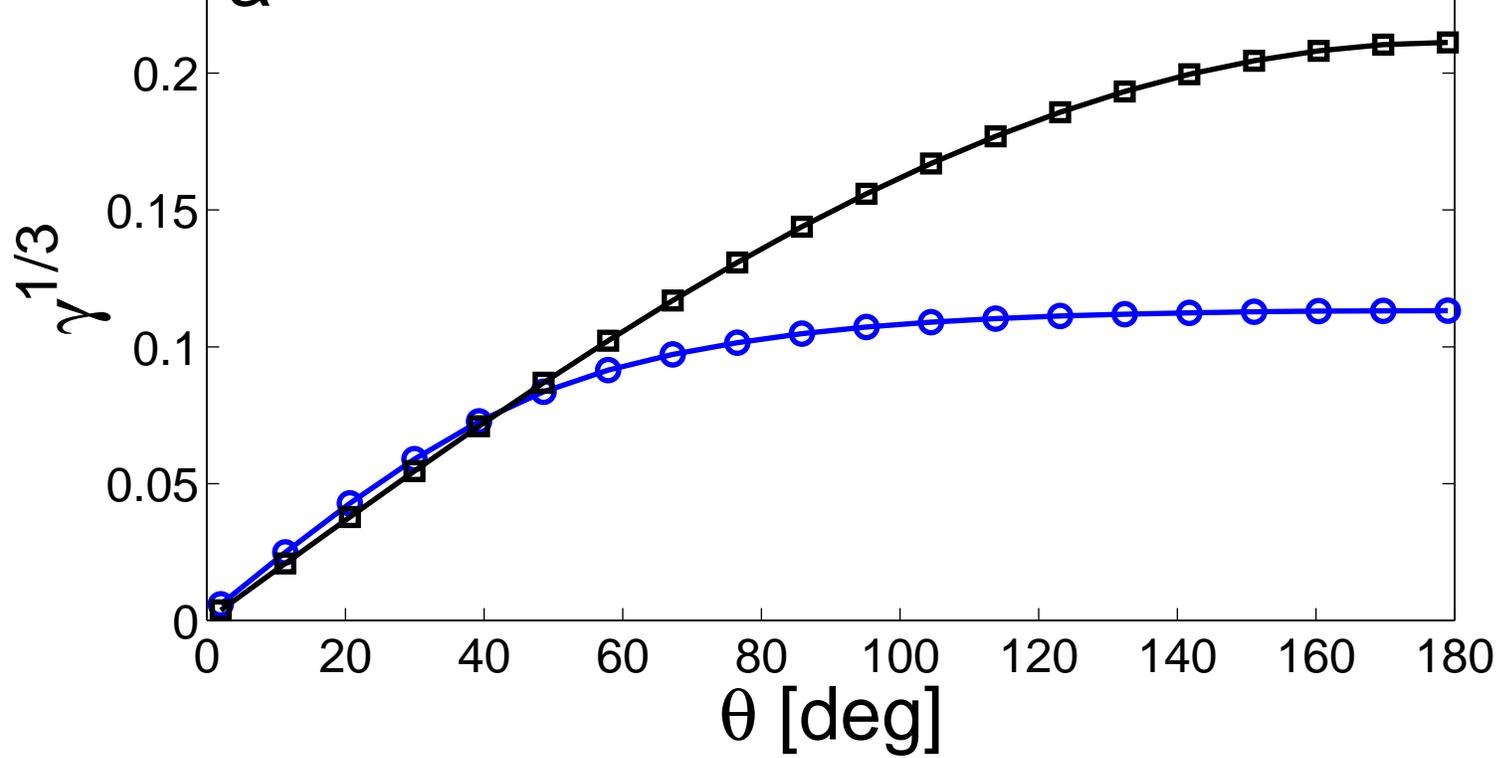
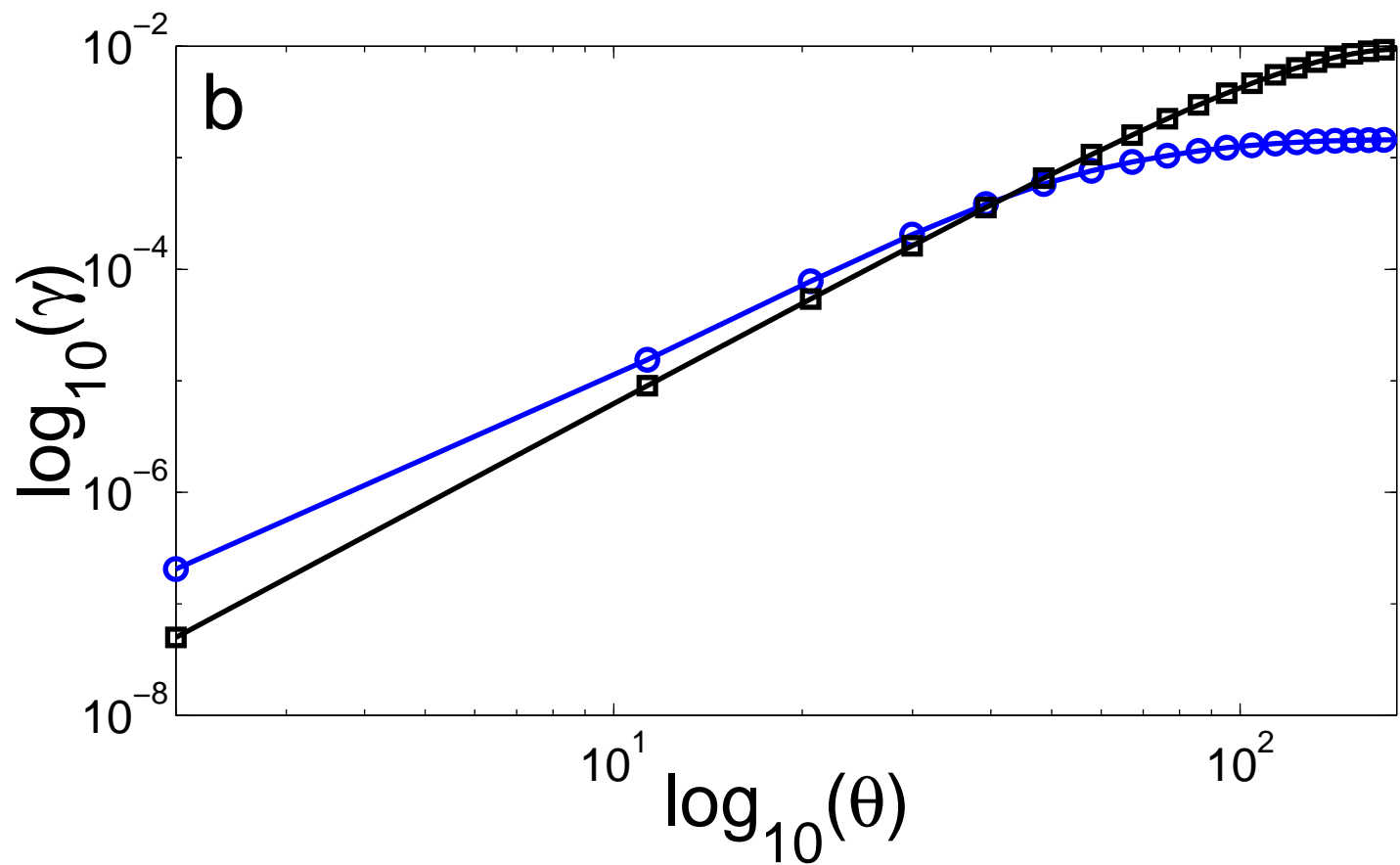
\item{\bf Fig.~8:}
In (a) the grain-boundary energy $\gamma$ of Fig.~6, is plotted as
$\gamma^{1/3}$ against the tilt angle $\theta$. The dashed curve with
rectangles is straight for small angles, depicting perfect $\gamma\sim\theta^3$
dependence. The curve marked with circles shows a power exponent 2.91. In (b)
the two line tension are plotted on a log-log plot. The results of Sec.
\ref{chevrons} show $\theta^3$ scaling, while the omegas show a smaller power
exponent. Possible small numerical errors in our scheme to find the wavevector
$k$ are important for small angles (the leftmost point being at
$\theta=2^\circ$), and this may easily change the line tension exponent from
the expected $3$ to $2.91$.

\item{\bf Fig.~9:}
The imaginary and real parts of the wave-vector $k$ in expression
(\ref{floquet form}) are shown in (a), for $\chi=1$.
The line with rectangles corresponds to the imaginary
part. Only the positive imaginary part is shown. The line with
circles corresponds to the real part of $k$. Dotted line shows $q_y$,
being the characteristic length scale in the $y$-direction.
Notice the splitting of the real
part of $k$ in the band gaps. An enlargement of the same
curve of the
imaginary part of $k$, for $17^{\circ}\leq\theta\leq50^{\circ}$
is shown in (b). The black lines indicates
integer multiples of $q_y$ in order
to show the regions of degeneracy.
Three band gaps are shown, around $18^{\circ}$,
$26^{\circ}$ and $42^{\circ}$. In these gaps ${\rm Im}(k)=imq_y$ for
some integer $m$.

\end{itemize}


\begin{thebibliography}{99}
\bibitem{chi_s}
Note that when $\chi_f$ takes the value $\chi_f^*$ so that $c_1=0$, one
might expect a continuous order-disorder transition in the symmetric A/B
system to a uniform phase, as occurs in ordinary phase
separation. However as will be shown, this transition is pre-empted by a
continuous transition to a modulated phase which occurs at $\chi=
\chi_f/\chi_f^*=3/4$. At this value, $c_1$ is still positive.
\end{thebibliography}
\end{document}